\documentclass[oldversion]{aa}
\usepackage{txfonts}
\usepackage{epsfig}
\usepackage{graphicx}
\usepackage{natbib}
\bibpunct{(}{)}{;}{a}{}{,}
\begin{document}

\titlerunning{Forming an Early O-type Star Through Gas Accretion?}
\authorrunning{Zapata, Palau, Ho, Schilke, Garrod, Rodr\'\i guez, and Menten}
\title{Forming an Early O-type Star Through Gas Accretion?}

\author{Luis A. Zapata\inst{1}, Aina Palau\inst{2}, Paul T. P. Ho\inst{3,4},
Peter Schilke\inst{1},  Robin T. Garrod\inst{1}, \\ 
Luis F. Rodr\'\i guez\inst{5}, and Karl Menten\inst{1}}

\institute{Max-Planck-Institut f\"{u}r Radioastronomie, Auf dem H\"ugel 69,
53121, Bonn, Germany
\and Laboratorio de Astrof\'\i sica Espacial y F\'\i sica Fundamental, Apartado 78
E-28691, Villanueva de la Ca\~{n}ada, Madrid, Spain.
\and Harvard-Smithsonian Center for Astrophysics, 60 Garden Street,
Cambridge, MA 02138, USA
\and Academia Sinica Institute of Astronomy and Astrophysics,
Taipei, Taiwan
\and CRyA, Universidad Nacional Aut\'onoma de M\'exico,
Apdo. Postal 3-72 (Xangari), 58089 Morelia, Michoac\'an, M\'exico}

\date{Received -- / Accepted --}

\offprints{Luis Zapata, \email{lzapata@mpifr-bonn.mpg.de}}

\abstract {We present high angular resolution ($\sim$ 3$''$) and
sensitive 1.3 mm continuum, cyanogen (CN)
and vinyl cyanide (C$_2$H$_3$CN)
line observations made with the Submillimeter Array (SMA) toward one of  most
highly obscured objects of the W51 IRS2 region, W51 North.  We find that the CN
line exhibits a pronounced inverse P-Cygni profile indicating that the molecular gas is infalling inwards
this object with a mass accretion rate between 4 and 7 $\times$ 10$^{-2}$ M$_\odot$ yr$^{-1}$.
The C$_2$H$_3$CN traces an east-west rotating
molecular envelope that surrounds either a single obscured (proto)star with a kinematic mass
of 40 M$_{\odot}$ or a small central cluster of B-type stars and that is associated with a compact high velocity
bipolar outflow traced by H$_2$O masers and SiO molecular emission. 
We thus confirm that the W51 North region is part of the growing list
of young massive star forming regions that have been associated with infalling
motions and with large mass accretion rates ($\sim$ 10$^{-2}$ -- 10$^{-4}$  ),  
strengthening the evidence for massive stars forming with 
very high accretion rates sufficient to quench the formation of an UCHII region. 
}


\maketitle

\section{Introduction}

One of the main questions related to star formation is whether
massive stars ($>$ 10 M$_{\odot}$) are formed through
gas accretion via a circumstellar disk/torus or whether other mechanims play a role.
It was believed that the powerful radiation fields and stellar winds produced at
the very beginning of nuclear burning will increasingly inhibit farther accretion of material thereby
limiting the maximum stellar mass to about $10$ M$_{\odot}$ \citep{Kahn1974, LarsonStarrfield1971, YorkeKruegel1977}.
Several theoretical models have since been proposed to solve this puzzle: the formation of massive stars through dense
disks with jets/outflows \citep{Nakano1989, JijinaAdams1996}, through merging of smaller stars \citep{Bonnelletal1998},
through turbulent accretion \citep{MckeeandTan2003},  through competitive accretion \citep{Bonnellandbate2001},
and through ionized accretion flows \citep{KetoWood2006}. However, due to the lack of good observational guidance,
these alternatives have remained controversial \citep{ZinneckerandYorke2007}.

With a total bolometric luminosity of about 3 $\times$ 10$^6$ L$_{\odot}$ the W51-IRS2 region is one of the most 
luminous massive star forming regions in our Galaxy,
\citep{EricksonTokunaga1980}. It is located 6--7 kpc away in  the Sagittarius spiral arm \citep{Genzeletal1981, Imaietal2002}.
We adopt here a distance to the W51 region of 7 kpc.
The W51 IRS2 region comprises a complex group of highly obscured young objects (with not mid-infrared counterparts, 
see please \citet{Kraemeretal2001,Okamotoetal2001})
called ``W51 North'' and ``W51d2'', and a cluster of massive strong infrared ZAMS
stars \citep{Kraemeretal2001,Okamotoetal2001, Lacyetal2007},
which the most prominent member being the source  ``IRS2d'' associated with an extended edge-brightened
cometary HII region called ``W51d'' \citep{Gaumeetal1993, Lacyetal2007}.

The W51 North object shows strong thermal dust emission at (sub)millimeter wavelengths,
molecular emission from a large ($\sim$ 4 $\times$ 10$^4$ AU) {\it hot core} at
an excitation temperature of 100-200 K \citep{Zhangetal1998},
and very faint centimeter free-free emission \citep{Gaumeetal1993, Zhangetal1998, Eisneretal2002},
suggesting that it is forming an extremely young massive star.
This object, in addition, contains a group of strong H$_2$O, OH and SiO masers that are located in
the center of this molecular and dusty structure (called {\it "The Dominant Center"})
\citep{Schnepsetal1981, GaumeMutel1987, Hasegawaetal1986, Ukitaetal1987, Moritaetal1992}.
Observations of the proper motions of the H$_2$O masers (which traces shocks in dust-laden gas close to
the exciting protostars, Elitzur 1992) revealed the presence of a compact ($\sim$ 7000 AU)
northwest-southeast high velocity ($>$ 100 km s$^{-1}$) outflow \citep{Schnepsetal1981,
Eisneretal2002, Imaietal2002}.
Moreover, high angular resolution observations showed
that the SiO masers seem to be tracing the innermost parts of this powerful outflow \citep{Eisneretal2002}.
Finally, this object has previously been identified  with
spectroscopic signatures of dynamical collapse using emission
from HCO$^{+}$ \citep{Rudolphetal1990}, and SO$_{2}$ \citep{Sollinsetal2004}.

Here we present 1.3 mm continuum, cyanogen
and vinyl cyanide line observations
toward the W51 North region made with the SMA.
We report the presence of molecular gas accretion onto a 40 M$_{\odot}$ 
(proto)star or a small central cluster of B stars located in the center of 
W51 North region, and with an accretion
rate between 4 and $7\times$ 10$^{-2}$ M$_\odot$ yr$^{-1}$ .

\begin{figure*}[ht]
\begin{center}
\includegraphics[scale=0.44]{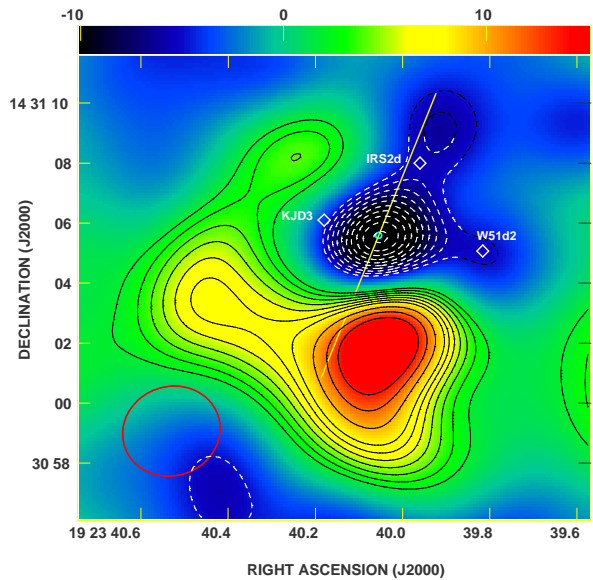}\vspace{-0.8cm} 
\caption{\scriptsize
Integrated molecular emission of the lines CN N=2-1, $J = 5/2-3/2, F = 5/2-3/2$ 
and  N=2-1, $J = 5/2-3/2, F =7/2-5/2$ observed with the SMA is shown as a color
image of the W51 IRS2 region. The contours are -14, -13, -12, -11, -10,
-9, -8, -7, -6, -5, -4, 4, 5, 6, 7, 8, 10, 11, 12, 13, and 14 times
970 mJy beam$^{-1}$, the rms noise of the image.
The integration is over a velocity range from 50 to 80 km s$^{-1}$.
The synthesized beam is 3.4 $''$ $\times$ 3.2$''$ with a P.A. = -87$^\circ$, and
is shown in bottom left corner.
The scale bar indicates the molecular line emission and absorption in Jy Beam$^{-1}$.
The white diamonds indicate the positions of the infrared sources KJD3 and IRS2d
\citep{Kraemeretal2001, Okamotoetal2001, Lacyetal2007} and the radio
source W51d2 \citep{Gaumeetal1993}.
The yellow line indicates the orientation of the SiO(5-4) high velocity
bipolar molecular outflow centered on the water masers (Zapata et al. in prep.).
The green circle indicates the position of the center of W51 North \citep{Schnepsetal1981}.}
\label{fig1}
\end{center}
\end{figure*}

\section{Observations}

The observations were made with the SMA\footnote{The Submillimeter Array (SMA)
is a joint project between the Smithsonian Astrophysical Observatory and
the Academia Sinica Institute of Astronomy and Astrophysics, and is
funded by the Smithsonian Institution and the Academia Sinica.}
during 2005 August 20. The SMA was in its compact configuration,
which includes 21 independent baselines ranging in projected length from 16 to 50 m.
The phase reference center of the observations was
R.A. = 19h23m43.80s, decl.= 14$^\circ$31$'$30.0$''$ (J2000.0).
The frequency was centered at 217.1049 GHz in the Lower Sideband (LSB),
while the Upper Sideband (USB) was centered at 228.1049 GHz.

A close blend of the CN N=2-1, $J = 5/2-3/2, F = 5/2-3/2$ and  N=2-1, $J = 5/2-3/2, 
F =7/2-5/2$ lines were detected in the USB. Their frequencies are
226.874166 and 226.874745 GHz, respectively.
Both lines have very similar intrinsic strengths and energies above the ground
state. The LSR velocity scale in this paper is given with respect to
the rest frequency of the former line. The velocity difference
corresponding to the frequency difference is 0.77 km s$^{-1}$. 
When, in this paper, we refer to ``the'' CN line, we mean this blend.
The C$_2$H$_3$CN $J_{K_a, K_c}=
23_{2,22}-22_{2,21}$ line was detected in the LSB at a frequency of
217.497585 GHz.

The full bandwidth of the SMA digital correlator is 4 GHz (2 GHz in
each side band).  The correlator was configured with spectral windows
(``chunks'') of 104 MHz each, with 128 channels distributed over each
spectral window, providing a resolution of 0.8125 MHz (1.1 km
s$^{-1}$) per channel.

The zenith opacity measured ($\tau_{230 GHz}$) with the NRAO tipping
radiometer located at the Caltech Submillimeter Observatory (close to
the SMA) varied during the night between 0.12 and 0.20, indicating
good weather conditions during the observations.  Phase and amplitude
calibrators were the quasars 1749+096 and 1741$-$038, with measured
flux densities and formal fitting errors of 2.08 $\pm$ 0.05 and
1.81 $\pm$ 0.05 Jy, respectively.  The uncertainty in the flux scale
is estimated to be 15--20$\%$, based on the SMA monitoring of quasars.
Observations of Uranus provided the absolute scale for the flux
density calibration.  Further technical descriptions of the SMA and
its calibration schemes are found in \citet{Hoetal2004}.

The data were calibrated using the IDL superset MIR, originally developed 
for the Owens Valley Radio Observatory \citep{Scovilleetal1993} and adapted for the 
SMA.\footnote{The MIR-IDL cookbook by C. Qi can be found at
http://cfa-www.harvard.edu/$\sim$cqi/mircook.html} 
The calibrated data were imaged and analyzed in the standard manner
using the MIRIAD and AIPS packages.  We used the ROBUST parameter of
the INVERT task set to $-2$, which corresponds to uniform weighting to
achieve the maximum angular resolution while sacrificing some
sensitivity.  The resulting image rms noise of line images was 30 mJy
beam$^{-1}$ for each channel at an angular resolution of 3.4$''$
$\times$ 3.2$''$ with a P.A. = -87$^\circ$.  The data were
self-calibrated in phase and amplitude using as a model the continuum
image.  The final images were shifted $\sim$ 1$''$ in right ascension
in order to be consistent with the positions of the molecular cores
associated with the W51 North and W51d2 better determined with the
Very Large Array observations of
\citet{Hoetal1983, Zhangetal1997}. This discrepancy is mainly caused by 
the baseline error, the finite S/N, and the atmospheric fluctuations 
in our millimeter wave observations.

\begin{figure}[ht]
\begin{center}
\includegraphics[scale=0.24, angle=0]{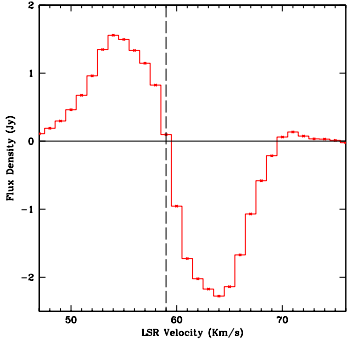}
\caption{\scriptsize  Spectrum of the CN N=2-1, $J = 5/2-3/2, F = 5/2-3/2$ and  N=2-1, $J = 5/2-3/2, 
F =7/2-5/2$  lines towards the center
of the W51 North object.  The spectral velocity resolution is 1.1 km s$^{-1}$.  
The dashed line indicates the systemic velocity (V$_{LSR}$=59 km s$^{-1}$). }
\label{fig2}
\end{center}
\end{figure}

\section{Results and Discussion}

\subsection{The Molecular and Millimeter Continuum Emission}

In Figure \ref{fig1}, we present the
integrated CN line emission image of the W51 IRS2
region, and the positions of the H$_2$O masers \citep{Imaietal2002},
the infrared sources W51 IRS2d and KJD3/OKYM1
\citep{Kraemeretal2001,Okamotoetal2001, Lacyetal2007}, and the radio
source W51d2 \citep{Gaumeetal1993}, all of them in the neighborhood of
the W51 North object.  Furthermore, we mark the position and
orientation of the strong high velocity molecular bipolar outflow traced by the 
SiO \textit{v}$ =0; J=5 - 4$ line found by Zapata et al. (in prep.), which is
centered on the cluster of masers.  In this image we can see two
components of the molecular line distribution, one in unresolved absorption
toward the W51 North object and the other also quite compact, but resolved
in emission surrounding this region.  
We interpret this appearance the
signature of molecular absorption against the strong compact
millimeter continuum source associated with W51 North and shown in
Figure 3.

Figure \ref{fig2} shows the spectrum of the CN
line towards the center of the W51 North
object. The line shows a pronounced inverse P Cygni profile.  The
emission feature appears at a V$_{LSR}$=55 km s$^{-1}$ and absorption
at V$_{LSR}$=65 km s$^{-1}$, which is consistent with the HCO$^+$
molecular observations of \citet{Rudolphetal1990}. Given that the
core's  systemic velocity is 59 km s$^{-1}$ \citep{Zhangetal1998},  
the location of the redshifted absorption projected against 
the bright continuum emission of the central
highly obscured object implies inward motions away from the observer.
The brightness temperature of the emission from the infalling material 
is lower than $\sim$ 10 K, the brightness temperature of the continuum 
emission from the core.  From Figure 6 of \citet{Rudolphetal1990} and
Figure 2 shown here, we estimate that the velocity of infall
($V_{infall}$) is about 4 km s$^{-1}$.  With this information and
taking the values of the density ($\rho$=2$\times$10$^{6}$ cm$^{-3}$),
linear radius (r=1.4 $\times$10$^4$ AU) reported for the compact
continuum source located in W51 North \citep{Zhangetal1997,
Zhangetal1998} and a radius of the hot core of 2 $\times$ 10$^4$ AU,
and following Beltran et al.  (2006), we calculate that the mass
infall rate (\.{M}=$4 \pi r^2 \rho V_{infall}$) is between 4 and $7
\times 10^{-2}$ M$_\odot$ yr$^{-1}$. This value  
have large uncertainties, due to the 
uncertainty on the density and on the radius at which V$_{infall}$ is 
measured.

The first moment map of the C$_2$H$_3$CN {\bf $J_{K_a, K_c}=
23_{2,22}-22_{2,21}$}  line is shown
in Figure 3. The emission is tracing an unresolved 
east-west rotating molecular ``envelope'' or ``core''
with a total velocity shift of 1.5 km s$^{-1}$ and a size of 4 $\times$ 10$^4$ AU.
Moreover, the integrated emission
from this molecule is well centered on the millimeter continuum source,
suggesting that this species is tracing high density gas close to the (proto)star  
(see Figure 3).
Finally, our bandpasses contained lines from other molecules 
(e.g. HCOOCH$_3$ and CH$_3$OH)
associated with W51 North. However, they were 
very much contaminated by
the emission from the hot molecular core associated with W51d2, not allowing us to search for 
similar east-west velocity gradients associated with these line molecular tracers.
The W51 North source
is associated with the extended {\it hot core} found by
\citet{Hoetal1983} and \citet{Zhangetal1998}, and it is centered on 
the cluster of masers as reported in other
observations e.g. NH$_3$; \citet{Hoetal1983,Eisneretal2002}, CH$_3$CN;
\citet{Zhangetal1998}, SO$_2$; \citet{Sollinsetal2004}.  If we assume
that the molecular gas is rotating as a rigid body (i.e. the dynamical
mass is M$_{dyn}$=v$^2$rsin$^2$({\it i})/G, where v is the rotation
velocity, r is the radius of the envelope, {\it i} is the inclination
angle of the envelope assumed to be 90$^\circ$ and G is the
gravitational constant), we estimate a mass for the central object(s) of
40 M$_\odot$.  This central object might be associated with a single central
O-type (proto)star or with a small group of B-type (proto)stars.
However, as there is a strong and compact bipolar outflow in the center of the core 
(see Figure 3), it seems to be dominated by one central massive star.

From Figure 3 and assuming that at a wavelength of 1.3 mm 
we are observing isothermal optically thin dust
emission with a dust mass opacity coefficient that varies 
with frequency as $\kappa \propto
\nu^{\beta}$, with $\beta=1$, (the size of the source suggest 
that we are observing emission from the envelope; hence, we adopt $\beta$=1, 
however, this value is uncertain, see \citet{Beckwithetal1990} that observe 
how this value varies in pre-main-sequence stars),
a gas-to-dust ratio of 100 (which may not be the most adequate to use for 
protostellar sources since erosion of the circumstellar envelope by 
photoevaporation from near OB stars may decrease 
the gas-to-dust ratio, see \citet{Williamsetal2005, ThroopBally2005}),
an adopted value of $\kappa_{1.3 mm}$ = 1.5 cm$^2$ g$^{-1}$ and a 
dust temperature value of  about 100 K (with uncertainties 
of a factor of about 1.5, \citet{Zhangetal1998}), we estimate an enclosed mass of the molecular core 
W51 North of 90 M$_\odot$, very close to the 100 M$_\odot$ estimated by \citet{Zhangetal1997}.
Due to the uncertainties referred to above, the values of the derived masses 
are good within a factor of 2.

\begin{figure}[ht]
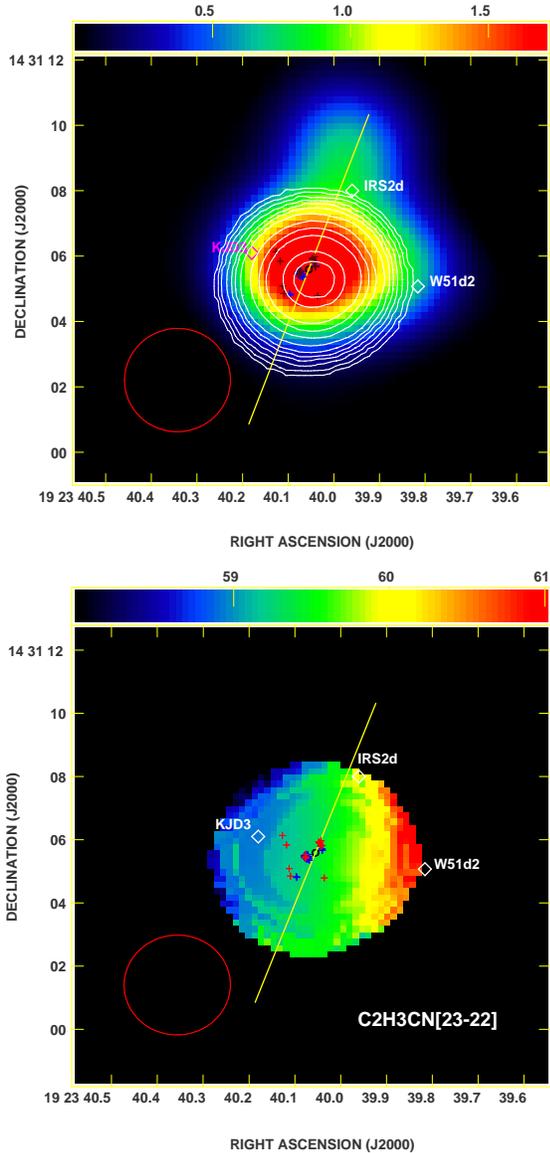

\begin{center}
\includegraphics[width=7.8cm, height=8.5cm]{f3.eps}\\ \vspace{-0.8cm} 
\includegraphics[width=7.83cm, height=8.9cm]{f4.eps}
\caption{\scriptsize
{\it \bf Upper:} SMA 1.3 mm continuum emission color image overlayed
with the  moment zero distribution of the C$_2$H$_3$CN line 
(pink contours) of
the W51 North region. The integration is over a velocity range of 55
to 65 km s$^{-1}$.  The synthesized beam with a FWHM 3.4 $''$ $\times$ 3.2$''$
and a P.A. of  $-87^\circ$ is shown in the bottom left corner.  The
contours are -4, 4, 8, 12, 16, 20, 24, 30, 40, 50, 60, 80, 90, 100,
120, 150, 170 and 200 times 40 mJy beam$^{-1}$ km s$^{-1}$, the rms
noise of the image.  The scale bar indicates the continuum peak flux
density in Jy beam$^{-1}$.  {\it \bf Lower:}  
First moment color image of the C$_2$H$_3$CN 
emission toward the W51 North region.  The
integration is over a velocity range of 55 to 65 km s$^{-1}$.  The
scale bar indicates the velocity shift in km s$^{-1}$.  The white
diamonds indicate the positions of the infrared sources KJD3 and IRS2d
\citep{Kraemeretal2001, Okamotoetal2001, Lacyetal2007}
and the radio source W51d2 \citep{Gaumeetal1993}.
The blue and red crosses indicate the position of the blue- and red-shifted
strong H$_2$0 masers spots, respectively, reported by \cite{Imaietal2002}.
Note that the central cluster of masers is
tracing a high velocity outflow with a northwest-southeast orientation
\citep{Imaietal2002, Eisneretal2002}.
The yellow line indicates the orientation of the SiO(5-4)  high velocity
bipolar outflow centered on the water masers (Zapata et al. in prep.).
The black circle indicates the position of the center of W51 North \citep{Schnepsetal1981}.}
\end{center}
\label{fig3}
\end{figure}

\subsection{Forming an Early O-type star in W51 North?}

The combined 1.3 mm continuum, C$_2$H$_3$CN and
CN data from W51 North suggest that this object is forming a massive O5-type
(proto)star in its center through molecular gas accretion and with a
very large accretion rate between 4 and $7\times$ 10$^{-2}$ M$_\odot$
yr$^{-1}$.  Moreover, the powerful compact bipolar high velocity ($>$ 100 km s$^{-1}$) 
outflow traced by the H$_2$O masers pinpoints the position of this putative 
central massive object.

Zapata et al. (in prep.) in addition found a compact high velocity  
SiO bipolar outflow with both unresolved-lobes spatially separating 
for less than one arcsecond and forming a P. A. of 150$^\circ$ $\pm$ 30$^\circ$.
This is centered at the position of the compact
H$_2$O outflow which has a P.A. of 110-145$^\circ$ \citep{Schnepsetal1981,
Imaietal2002, Eisneretal2002}.  We propose that these two outflows
might be manifestations of a single powerful outflow.  The masers appear to be
tracing the innermost regions of it  (as observed in other outflows,
e.g. IRAS 20126+4104, Moscadelli et al. 2000), while the SiO is
tracing the more extended shocked molecular gas. From this point of
view, the southeastern cluster of H$_2$O masers (see Figure 3)
maybe tracing another older ejected bow-shock. However, higher angular
resolution SiO molecular observations are neccesary to confirm this
picture.  It is interesting to note that the P.A. of the velocity gradient
across the C$_2$H$_3$CN envelope ($\sim$ 90$^\circ$) is not exactly 
perpendicular to the orientation of the molecular outflow 
(P.A. $\sim$ 140$^\circ$), as might be expected. This may indicate that outflow could be 
precessing due the presence of binary system or 
that the C$_2$H$_3$CN  emission is
contaminated by the outflow. 
This physical phenomenon of precessing outflows has been reported
in other outflows: IRAS 20126+4104
\citep{Shepherdetal2000}, L115{\bf 7} \citep{Bachilleretal2001}, 
and NGC7538IRS1 \citep{Krausetal2006}.

At present, there is a list of early massive (proto)stars that have
been associated with possible infalling motions and with large mass
accretion rates, e.g.  W51e2: a gas mass of 200 M$_{\odot}$ and an
accretion rate of 10$^{-3}$ M$_\odot$ yr$^{-1}$ \citep{Zhangetal1997,Hoandyoung1996};
NGC7538-IRS9: a gas mass of 100-300 M$_{\odot}$ and an accretion rate
of 10$^{-3}$ M$_\odot$ yr$^{-1}$ \citep{Sandelletal2005}; G24.78+0.08: 
a (proto)stellar mass of $\sim$ 20 M$_\odot$ and an accretion rate
between 10$^{-2}$ to 10$^{-4}$ M$_\odot$ yr$^{-1}$
\citep{Beltranetal2006}; IRAS 16547-4247: associated with an O-type
(proto)star and an accretion rate of about 10$^{-2}$ M$_\odot$
yr$^{-1}$ \citep{Garayetal2007} and W51 North, a (proto)stellar mass
of $\sim$ 40 M$_\odot$ and an accretion rate between 4-7 $\times$
10$^{-2}$ M$_\odot$ yr$^{-1}$  (these results). This suggests that
the very massive stars (O-type) might form starting with very high
accretion rates, sufficient to quench the formation of an UCHII
region.  We note that this hypothesis has already been proposed by
recent numerical simulations \citep{Banerjeeetal2007}.

We thank the anonymous referee for many valuable suggestions.
R.G. is grateful to the {\it Alexander von Humboldt Foundation} 
for a Humboldt Research Fellowship.

\bibliographystyle{aa}
\bibliography{zapata}

\end{document}